\documentclass{aa} 
\usepackage{graphicx}
\usepackage{txfonts}
\usepackage{amsmath}

\usepackage{amssymb}
\usepackage{natbib}
\usepackage[utf8]{inputenc}
\usepackage[T1]{fontenc}
\usepackage{lmodern}
\usepackage[dvipsnames]{xcolor}
\usepackage[colorlinks=true,allcolors=blue]{hyperref}
\usepackage{orcidlink} 

\newcommand{\uint}{u_\mathrm{int}}
\newcommand{\pgas}{p_\mathrm{gas}}
\newcommand{\pmag}{p_\mathrm{mag}}
\newcommand{\prad}{p_\mathrm{rad}}

\defcitealias{lancova_new_2025}{\texttt{L26}}
\defcitealias{2013MNRAS.428.2255P}{\texttt{P13}}
\defcitealias{2025MNRAS.542..377R}{\texttt{R25}}

\begin{document}
\title
{
 Universal behaviour of $\alpha$-viscosity in black hole accretion discs 
}
%
\author{          Marek A. Abramowicz\inst{1,2,3}\thanks{\email{marek.abramowicz@physics.gu.se}}\orcidlink{0000-0003-0067-5895}
\and              Axel Brandenburg\inst{4,5}
\orcidlink{0000-0002-7304-021X}
\and              Ji{\v r}{\'\i} Hor{\'a}k\inst{6}
\orcidlink{0000-0002-7635-4839}
\and              Debora Lan\v{c}ov\'a\inst{2,3}\thanks{\email{debora.lancova@physics.cz}}\orcidlink{0000-0003-0826-9787}
\and \\         John C. Miller\inst{7,8}
\orcidlink{0000-0002-1708-4330}
\and              Ewa Szuszkiewicz\inst{9}
\orcidlink{0000-0002-7881-2805}
\and              Maciek Wielgus\inst{10}
\orcidlink{0000-0002-8635-4242}
}
\institute{    Physics Department, Gothenburg University,              
               SE-412-96 G{\"o}teborg, Sweden                       
\and           Nicolaus Copernicus Astronomical Center, Polish Academy of Sciences, Bartycka 18, PL-00-716 Warszawa, Poland        
\and           Research Center for Computational Physics and Data Processing, Institute of Physics,                                   
               Silesian University in Opava, Bezru{\v c}ovo n{\'a}m.   %
               13, CZ-746 01 Opava, Czech Republic                  
\and           Nordita,                                                
               Hannes Alfvéns v\"ag 12, SE-106 91 Stockholm, Sweden 
\and           The Oskar Klein Centre, Department of Astronomy,        
               Stockholm University, AlbaNova, 10691 Stockholm, Sweden %
\and           Astronomical Institute of the Academy of Sciences,      
               Bo{\v c}ni II 1401/1a,                                  %
               CZ-141 31 Praha 4, Czech Republic                    
\and           Department of Physics (Astrophysics),               
               University of Oxford, Keble Road, Oxford, OX1 3RH,      %
                                                   United Kingdom   
\and           International School for Advanced Studies, SISSA,       
              Via Bonomea 265, 34136 Trieste, Italy      %
\and           Institute of Physics and CASA*, University of Szczecin, ul. Wielkopolska 15, PL-70-451 Szczecin, Poland  
\and          Instituto de Astrofísica de Andalucía-CSIC, Glorieta de 
              la Astronomía s/n, E-18008 Granada, Spain              
}
   \date{Received ????; accepted ????}
\abstract{
The Shakura-Sunyaev $\alpha$-viscosity coefficient, defined as the ratio of total stress to total pressure, $\alpha= \mathbb{T}/p$, began to play an important role in the development of accretion disc theory in the early 1970s. The origin of the turbulence that causes the stress $\mathbb{T}$ was unknown at that time; Shakura and Sunyaev assumed $\alpha=$\,const. Today we know that this was not quite realistic --- modern general relativistic magneto-hydrodynamic simulations (GRMHD) of black hole accretion discs have revealed that $\alpha$ changes by about an order of magnitude within the disc, being smaller far away from the black hole and larger in the plunging region close in, and it has been found that the behaviour of $\alpha$ reflects some underlying, fundamental properties of the stress $\mathbb{T}$. In particular, it has been argued by several authors, that $\mathbb{T}$ must be zero at the black hole horizon. We note that the stress calculated in three independent GRMHD simulations of accretion discs around non-rotating black holes, made by a variety of authors (including ourselves), each has its prominent maximum close to the location of the circular photon orbit. We propose a formula that accurately describes this ``universal'' behaviour of $\alpha$ in terms of the ``gyration radius'', a physical characteristic of rotation well known in Newtonian dynamics and in the black hole case uniquely defined by the Kerr space-time geometry. Analytic and semi-analytic models of black hole accretion discs provide an invaluable insight into fundamental physics, and the GRMHD simulations do not aspire to replace them. Rather, simulations could help to improve analytic models by making them more realistic. For example, our $\alpha$-formula, deduced from the GRMHD simulations, may be useful in the construction of improved versions of thin and slim disc models.}

%
%
\authorrunning{Abramowicz, et al.}\titlerunning{A universal $\alpha$ prescription}
%
\maketitle
\setcounter{equation}{0}
\setcounter{figure}{0}
\setcounter{section}{0}
\nolinenumbers
\section{Motivation}
\label{sec:Motivation}  

The classic, analytic model by \citet{Shakura1973} is today still one of the pillars on which our understanding of the fundamental physics governing black hole accretion discs rests. The remarkable practical usefulness of the model results from its sharp clarity that was achieved by several brilliant assumptions that have been made. Perhaps the most important of them is the treatment of the ``viscous'' stress $\mathbb{T}$ that powers the accretion disc's radiation by enhancing angular momentum transport and dissipating orbital energy of matter into heat. Directly from the fundamental conservation laws, \citet{Shakura1973} derived equations for the radial fluxes of mass $\dot{M}$, angular momentum $\dot{J}$, and energy $\dot{E}$ in accretion discs:
\begin{align}
\dot{M} &= \int^{+H}_{-H} 2\pi r \, \rho v \, dz, \label{eqn:mass-flux}   \\          
\dot{J} &= \dot{M}\,j + \mathbb{T}, \quad                          
\dot{E} = \dot{M}\,e + \Omega \mathbb{T}. 
\label{eqn:momentum-energy-flux} 
\end{align}
Here $\rho$ is the mass density, $v$ is the radial velocity of the matter, integration goes through the disc vertical thickness\footnote{\label{foonote-01}For simplicity we consider here the Schwarzschild black hole and use spherical coordinates $\{t, r, \theta, \phi\}$ and metric signature $(+,-,-,-)$, in which $g_{tt} = c^2(1 - r_\mathrm{H}/r)$ and $g_{\phi\phi} =- \sin^2 (\theta)\,r^2$ with $r_\mathrm{H} = 2\,\mathrm{G}M/c^2$ being the horizon location and $M$ being the mass of the black hole. The Kerr black hole formulae are given in the Appendix \ref{app:Kerr-formulae}.}
\begin{equation} 
z=H(r), \quad z = r\cos(\theta),    
\label{eqn:vertical-thickness}
\end{equation}    
$j$ and $e$ are specific (per mass) angular momentum and energy, and $\Omega$ is the angular velocity. The difference $\Delta\dot{X}$ in a flux $\dot{X}$ between its radial locations $r_1$ and $r_2$ equals for the three fluxes (\ref{eqn:mass-flux}), (\ref{eqn:momentum-energy-flux}),  
\begin{equation} 
\Delta \dot{M} = 0, \quad \Delta \dot{J} =0, \quad \Delta \dot{E} = \int^{r_1}_{r_2} \, F \, 2\pi r \,dr.         
\label{eqn:conservation-mass-momentum}
\end{equation}    
The radial flux of energy is not constant because energy is also partially radiated from the disc; $F$ denotes the flux of energy that is locally radiated. From equations (\ref{eqn:mass-flux})--(\ref{eqn:conservation-mass-momentum}), \citet{Shakura1973} derived that the flux is given by
\begin{equation}
F = \frac{\dot{M}}{4\pi r}\left( \frac{d\Omega}{dr} \right) \left[ j\left(r\right) -{j\left(r_0\right)} \right],
\label{eqn:flux-equation}
\end{equation}
where $j(r)$ is the radial angular momentum distribution in the disc and $r_0$ is the radial location of the place
where the stress is zero, $\mathbb{T}(r_0)=0$. Note that 
\begin{equation}
j = \Omega \tilde{r}^2 
\quad \tilde{r} =\left( - \frac{g_{\phi \phi}}{g_{tt}}\right) ^{1/2}=\mbox{gyration radius}.
\label{eqn:omega-j}
\end{equation}
Knowing $r_0$ and $j(r)$, and therefore also $\Omega(r)$, suffices to calculate the local radiation flux from equation (\ref{eqn:flux-equation}).  

For a razor thin disc $h=H/r \ll 1$, with accuracy to linear order terms ${\cal O}^1(h)$, one may assume
\begin{align}
j(r) = \left\{ 
\begin{array}{lr}
        j_\mathrm{K}(r) = \tilde{r}^2 \Omega_\mathrm{K}(r)
                                                               & \text{~if~} r > r_0 \\
        j_\mathrm{K}(r_0) = \mbox{const}                                & \text{~if~} r < r_0         
        \end{array}  
  \label{eqn:Keplerian-momentum-ISCO}                                                \right\}
  \end{align}
where $\Omega_\mathrm{K}(r)$ is the Keplerian angular velocity and $r_0$ is the location of the innermost stable circular orbit (ISCO). Because  $\Omega_\mathrm{K}(r)$ and the location of the ISCO are known analytically in the Kerr geometry, 
from (\ref{eqn:Keplerian-momentum-ISCO}) one can derive the formula for the radiation flux emitted in an explicit (and very simple) form. Its Newtonian version reads
\begin{equation}
F(r) = \frac{3\mathrm{G}M{\dot{M}}}{{8\pi r^3}} \left[ 1 - \sqrt{\frac{r_0}{r}} \right], 
\quad r_0 = \frac{6\mathrm{G}M}{\mathrm{c}^2}.
\label{eqn:radiation-flux-famous}
\end{equation}
Formulae (\ref{eqn:flux-equation}) and (\ref{eqn:radiation-flux-famous}) are undoubtedly the most famous and useful in accretion disc theory. They have been used in Newtonian, Schwarzschild, and Kerr versions by numerous authors to calculate the observational appearance of accretion discs \citep[e.g.,][]{1973blho.conf..343N}. They do not depend on the stress $\mathbb{T}$ and this was an enormous benefit in the early days of the development of the theory, when the nature of the stress was unknown. Because the stress enters a few other equations on which their model rests, \citet{Shakura1973} have adopted an ansatz known as the Shakura-Sunyaev viscosity prescription, which states that the stress is proportional to the total pressure, 
\begin{equation}
\mathbb{T} = \alpha p, \quad \alpha = \mbox{const}.
\label{eqn:Shakura-Sunyaev-prescription}
\end{equation}
There is a problem here. The formula (\ref{eqn:radiation-flux-famous}) is valid only in the limit of a razor thin 
disc $h \rightarrow 0$, when only the lowest order ($\sim {\cal O}^1[h]$) terms are taken into account. When the disc luminosity is greater than about $\sim 0.3\,L_\mathrm{Edd}$, where $L_\mathrm{Edd}$ is the Eddington luminosity\footnote{and the corresponding mass accretion rate is $\dot{M}_\mathrm{Edd} = L_\mathrm{Edd}/(\eta_\mathrm{NT}\mathrm{c}^2)$, where $\eta_\mathrm{NT}$ is the \citet{1973blho.conf..343N} accretion efficiency.}, several important physical effects, in particular radial advection of heat described by the ${\cal{O}}^2[h]$ terms, must be taken into account because they {\it qualitatively} change the physics of accretion. In particular, neither $j(r) = j_\mathrm{K}(r)$ nor $\mathbb{T}(r_\mathrm{ISCO})=0$ are valid.

The slim disc model \citep{abramowiczSlimAccretionDisks1988b}, comprehensively discussed by \cite{2009ApJS..183..171S}, incorporates all of the ${\cal O}^2[h]$ terms that are missed in the thin disc model. In slim discs the stress is zero not at the ISCO, but at the horizon $r_\mathrm{H}$ and, correspondingly, the constant $j_0 = j(r_0)$ that appears in (\ref{eqn:flux-equation}) does not equal $j_\mathrm{K}(r_\mathrm{ISCO})$ as in thin discs, but instead is an eigenvalue of the problem, not known a priori. It cannot be assumed, but should be self-consistently calculated from the (extra) condition of regularity at the sonic radius $r_\mathrm{S}$. The formula (\ref{eqn:flux-equation}) only seemingly does not depend on the stress. In fact, when the terms ${\cal O}^2[h]$ are included, it depends strongly through the angular momentum distribution $j(r)$, which is directly governed by the stress. For this reason, in the plunging region of slim discs, that is located between the sonic radius and the horizon, $r_\mathrm{H} < r < r_\mathrm{S}$, the angular momentum is not constant and, correspondingly, the radiation flux $F(r)$ is not zero but may be quite large as calculations by \cite{2009ApJS..183..171S} have clearly demonstrated.

Obviously, the exact knowledge of the stress in the plunging region is important for calculating the radiation flux $F(r)$. Because slim disc models use the classic Shakura-Sunyaev ansatz for the stress (\ref{eqn:Shakura-Sunyaev-prescription}), the currently used slim disc values of $F(r)$ are not realistic. A more sophisticated description of the stress in slim discs is needed.

The importance of the plunging region emission has been recognised by many authors in both analytical and numerical studies \citep{krolikWhereInnerEdge2002, shafeeThreeDimensionalSimulationsMagnetized2008a, 2010MNRAS.408..752P, moralesteixeiraGeneralRelativisticRadiation2018, 2022MNRAS.515..775H, mummeryAccretionInnermostStable2023}. It was also demonstrated that the plunging region emission is important for the spin estimation in black hole binaries \citep[e.g.,][]{reynoldsObservationalConstraintsBlack2021,2022MNRAS.514..780W,mummeryRapidBlackHole2025,2025ApJ...980..203D,zdziarskiSpinsBlackHoles2026}. In those works, when a viscosity prescription is adopted, $\alpha$ is assumed to be constant throughout the disc, including the plunging region. Our formula extends this prescription by adopting a more physically motivated radial dependence of $\alpha$, with vanishing stress at the black hole horizon (as in slim discs) and with a maximum at the photon orbit, which more accurately reflects the behaviour of the accreting gas and magnetic fields in the vicinity of the event horizon.

\section{Our $\alpha$-prescription formula}
\label{sec:Our-alpha-formula}
\begin{figure} [!h]
\begin{center}
\includegraphics[width=0.98\columnwidth]{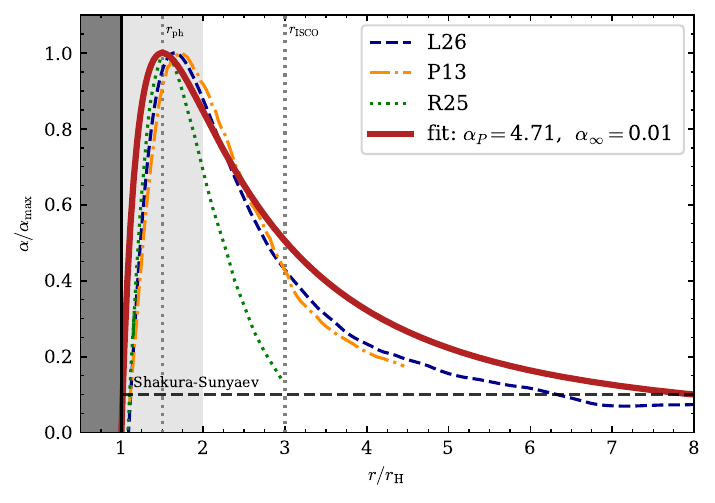}
\caption
{
The $\alpha$-viscosity coefficient calculated in GRMHD simulations by 
\citetalias{lancova_new_2025},
\cite{2013MNRAS.428.2255P}, denoted by \citetalias{2013MNRAS.428.2255P}, and 
\cite{2025MNRAS.542..377R}, denoted by \citetalias{2025MNRAS.542..377R},  all for accretion discs around non-rotating black holes, fitted to our $\alpha$-prescription formula (\ref{eqn:whole-domain}). Three characteristic features are visible in all of the simulations: [1] $\alpha=0$ at the horizon $r=r_{\mathrm H}$, [2] $\alpha$ has a maximum very close to the location of the circular light ray at $r_{\mathrm{ph}}=(3/2) \, r_{\mathrm H}$, [3] for large radii $r \gg r_{\mathrm H}$, $\alpha$ is much smaller than in the plunging region. The plunging region is located between the sonic radius $r_{\mathrm S}$ and the horizon $r_{\mathrm H}$ and is highlighted by the grey colour ($r_{\mathrm S}$ is taken from \citetalias{lancova_new_2025}). All lines are normalised by their respective maxima: $\alpha_{\mathrm{max}} = 0.8$ for \citetalias{lancova_new_2025}, $0.3$ for \citetalias{2013MNRAS.428.2255P}, and $0.3$ for \citetalias{2025MNRAS.542..377R}. 
Since \citetalias{2025MNRAS.542..377R} uses only the Maxwell stress in the local fluid frame (see Section \ref{sec:Stress-simulations}), their $\alpha$ definition is mostly valid in the highly magnetised plunging region and we therefore only show R25 data up to $r_\mathrm{ISCO}$.
} 
\label{fig:universal-stress}
\end{center}
\end{figure}

We suggest that the Shakura-Sunyaev constant $\alpha$-prescription formula (\ref{eqn:Shakura-Sunyaev-prescription}) should be replaced by\footnote{In the case of accretion discs around a non-rotating Schwarzschild black hole; see footnote \ref{foonote-01}.}
\begin{equation}
\alpha = \alpha(r) = \left[\alpha_P\left(\frac{r_{\mathrm{H}}}{r}\right)^2 + \alpha_{\infty} \right] \left(1 - \frac{r_{\mathrm{H}}}{r} \right).  
\label{eqn:whole-domain-coordinates}
\end{equation}
By $\alpha_P$ and $\alpha_{\infty}$ we denote two phenomenological constants: $\alpha_P=4.71$ and $\alpha_\infty=0.01$ give the best fit to the simulations by \citet{lancova_new_2025} denoted as \citetalias{lancova_new_2025} in this work; see Figure \ref{fig:universal-stress}.

The coordinate-independent version of (\ref{eqn:whole-domain-coordinates}), written with the help of the Killing vector $\eta^\nu$ (that describes the Schwarzschild metric time symmetry) and the Killing vector $\xi^\nu$ (axial symmetry) is\footnote{In Appendix \ref{app:Kerr-formulae}, we give the Kerr metric formulae.}   
\begin{equation}
\alpha = \alpha_P \left(\frac{r_{\mathrm{H}}}{{\tilde r}}\right)^2 + \alpha_{\infty} \left(\eta\eta \right).  
\label{eqn:whole-domain}
\end{equation}
We use the notation $(\xi\xi) = \xi^\mu \xi^\nu g_{\mu \nu}$, and likewise for $(\eta\eta)$.
The ``gyration radius'' $\tilde{r}$ is a key element in the arguments presented in this paper,
\begin{equation}
\tilde{r} = \left[-\frac{(\xi \xi)}{(\eta \eta)}\right]^{1/2}.
\label{eqn:gyration-radius}
\end{equation}
In Newtonian dynamics for a single particle orbiting a circle,
the gyration radius is just the radius of the orbit, whilst for rotating solids, the gyration radius relates to the moments of inertia. \cite{1993PhRvD..47.1440A} discussed in detail its physical meaning in Einstein's General Relativity. As in Newtonian dynamics, in Einstein's relativity for a~particle with four-velocity $u^\nu$ in the Schwarzschild space-time, the specific angular momentum $j=-u_\phi/u_t$ and the angular velocity $\Omega=u^\phi/u^t=d\phi/dt$ are related by equation (\ref{eqn:omega-j}).

The first term in (\ref{eqn:whole-domain}) dominates near to the black hole and far away it tends to zero. The second term dominates far away; it is introduced in order to reflect the asymptotic behaviour of $\alpha$ at large radii.
\section{Stress calculated from GRMHD simulations}
\label{sec:Stress-simulations}
%
%
\subsection{General remarks}
\label{sec:general-remarks}
In the GRMHD simulations of black hole accretion discs one usually assumes that the total stress-energy tensor is the sum of the Reynolds and Maxwell components,\footnote{We set here $\mathrm{G} = \mathrm{c} = M = 1$ for simplicity.}
\begin{align}
T_{\mu\nu}^\mathrm{Tot} &= T_{\mu\nu}^{\mathrm{Rey}} + T_{\mu\nu}^{\mathrm{Max}},                        \label{eqn:total-tensor} \\
T_{\mu\nu}^\mathrm{Rey} &= \ \varepsilon \, u_\mu u_\nu + \pgas g_{\mu\nu}, \label{eqn:Rey-tensor}\\
T_{\mu\nu}^\mathrm{Max} &= \, b^2 u_\mu u_\nu + \frac{1}{2}b^2g_{\mu\nu} - b_\mu b_\nu.                  \label{eqn:Max-tensor}
\end{align}
\noindent The four-velocity of matter is
\begin{equation}
u^\mu = \frac{dx^\mu}{ds} = \left\{ u^t, u^r, u^\theta, u^\phi \right\}
\label{eqn:four-velocity}
\end{equation}
and $\varepsilon = \rho + u_{\mathrm{int}} + \pgas$ is the total mass-energy density, with $\rho$, $\uint$, and $\pgas$ being the matter density, internal energy, and pressure, and  $b^\mu$ is the magnetic field four-vector. In the simulations, one solves the conservation equation 
\begin{equation}
\nabla^\mu\, T_{\mu\nu}^\mathrm{Tot}=0 
\label{eqn:conservation-equation-basic}
\end{equation}
together with the material equations (equation of state), the induction equation for the magnetic field, and proper boundary conditions on the chosen metric background that defines the covariant derivative $\nabla_\mu$. The continuity equation
\begin{equation}
\nabla_\mu (u^\mu \rho) = 0
\label{eqn:continuity}
\end{equation}
assures that matter is neither created nor destroyed during accretion.

The three simulations, \citetalias{lancova_new_2025}, \citetalias{2013MNRAS.428.2255P}, and \citetalias{2025MNRAS.542..377R}, mentioned in the present paper, were performed for accretion onto a non-rotating black hole described by the Schwarzschild metric, with $M=10\,\mathrm{M}_\odot$ for \citetalias{lancova_new_2025} (the other two used non-radiative GRMHD, which is scale-free). All are three-dimensional; \citetalias{2013MNRAS.428.2255P} and \citetalias{2025MNRAS.542..377R} use an initial torus threaded by magnetic loops with opposite polarity, and the same prescription for cooling, substituting the radiative losses \citep{2010MNRAS.408..752P}. Unfortunately, both of these simulations are rather short, with a total duration about $20\,000\,\mathrm{G}M/\mathrm{c}^3$, and the time-averaged properties, including $\alpha$, are only averaged over  $5\,000\,\mathrm{G}M/\mathrm{c}^3$ for \citetalias{2025MNRAS.542..377R} and  $13\,000\,\mathrm{G}M/\mathrm{c}^3$ for \citetalias{2013MNRAS.428.2255P}. During this short duration, the mass accretion rate is barely stable inside the averaging window. \citetalias{lancova_new_2025}, on the other hand, is evolved for more than  $80\,000\,\mathrm{G}M/\mathrm{c}^3$ and averaged over the last $17\,000\,\mathrm{G}M/\mathrm{c}^3$, with a stable mass accretion rate. This mainly affects the radial extent of the converged region of the simulations, which in the case of the short simulations covers only a few units of $r_{\mathrm{H}}$ above the radius $r_{\mathrm{ISCO}}$.

In contrast to the other two simulations, the puffy disc (\citetalias{lancova_new_2025}) uses \textit{radiative} GRMHD, where the gas conservatively exchanges energy and momentum with a frequency-integrated radiation field, a significantly more realistic approach. The non-radiative simulations use a cooling function to keep the accretion disc geometrically thin; thus, they omit radiation pressure, a crucial component for modelling the disc, and make a strong assumption about its structure.

The behaviour of the stress depends on local and quasi-local effects of the turbulence driven by the magnetorotational instability \citep[MRI;][]{Balbus1991} and on the global constraints given by the black hole geometry. It is natural to expect that the local effects are best described in the reference frame comoving with matter $e^\mu_{\,\,\,(k)}$  and the global constraints in the static reference frame attached to the space-time geometry $\tau^\mu_{\,\,\,[k]}$ \citep[see ][for the full formulation of this tetrad]{2011MNRAS.414.1183K}. The symbols $e^i_{\,\,\,(k)}$ and $\tau^i_{\,\,\,[k]}$ denote an orthonormal base of four-vectors:
\begin{equation}
e^\mu_{\,\,\,(t)} = \langle u^\mu \rangle_t, \quad \tau^\mu_{\,\,\,[t]} = n^\mu,
\label{eqn:tetrads}
\end{equation}
where $\langle u^\mu \rangle_t$ is time-averaged matter four-velocity (turbulence is smoothed out in time) and $ n^\mu=\eta^\mu [(\eta \eta)]^{-1/2}$ is the four-velocity of the static observer (in the Schwarzschild case). 
An invariant formula for the stress adopted in \citetalias{lancova_new_2025} and \citetalias{2013MNRAS.428.2255P} reads
\begin{equation}
\mathbb{T} = \langle T_{\mu \nu}  e^\mu_{\,\,\,(r)}\, e^\nu_{\,\,\,(\phi)} \rangle_{t,\phi} = \langle T_{(r) (\phi)}\rangle_{t,\phi} , 
\label{eqn:invariant-stress}
\end{equation}
where the angle-bracket subscripts label the averaging over time ($t$) and/or azimuth ($\phi$).

The physical meaning of this formula is explained in \citetalias{2013MNRAS.428.2255P}, including the effects of different ways of averaging the simulation. This approach provides a frame of a quasi-stationary flow (mean comoving frame), allowing us to study the stress in a way similar to how it was defined in the \citet{Shakura1973} disc model. 
\citetalias{2025MNRAS.542..377R}, on the other hand, calculates $\alpha$ using only the Maxwell stress in the local frame, defined by the instantaneous fluid velocity, which omits the Reynolds stress completely and only captures the instantaneous state of the fluid, rather than the quasi-stationary one.
%
%
\subsection{The puffy discs}
\label{sec:puffy-discs}

\begin{figure*}
\includegraphics[width=0.98\textwidth]{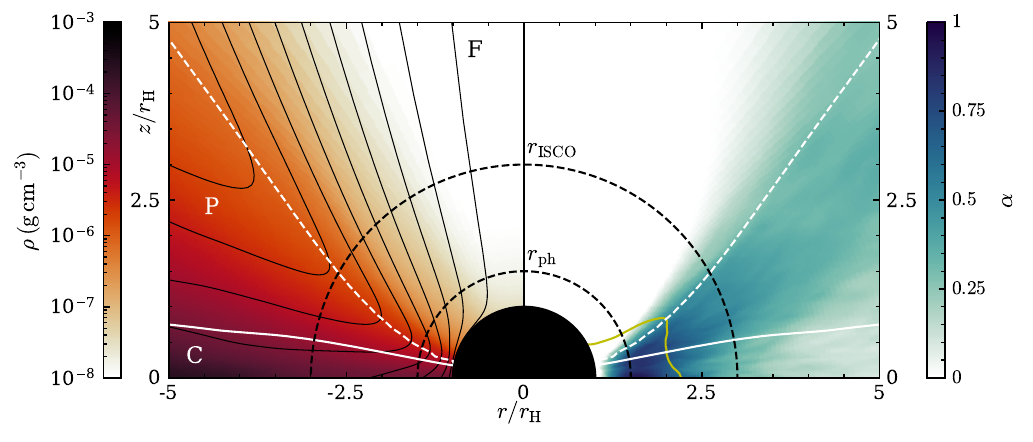}
\caption{
The structure of a puffy disc consists of three regions: C: dense core, P: puffy region, and F: funnel. \textit{Left:} The colour map of gas density with the magnetic field lines. \textit{Right}: Viscous $\alpha$ colour map, and the magnetosonic surface (yellow contour). In both panels, the white dashed line indicates the photosphere, and the full line indicates the density scale height. }
\label{fig:puffy}
\end{figure*}

The puffy accretion discs are mildly sub- and around-Eddington, radiation pressure-dominated discs around a~stellar-mass non-rotating black hole, modelled using three-dimensional global radiative GRMHD simulations \citepalias[\citealt{2019ApJ...884L..37L,2016MNRAS.459.4397S};][]{lancova_new_2025}. The simulations were performed using the \texttt{KORAL} code \citep{2013MNRAS.429.3533S}, and were designed so that the accretion disc self-consistently develops from a~dense torus acting as a source of matter far from the central black hole. Here, we used data from a simulation that reached the mass accretion rate $\dot{m}=0.6\,\dot{M}_\mathrm{Edd}$. The numerical methods and details of the puffy disc model can be found in \citetalias{lancova_new_2025}.

The puffy disc is stabilised against thermal instabilities, which are known to be critical in analytical radiation-pressure-dominated accretion discs,  by a vertical net flux of magnetic field \citep{2016MNRAS.459.4397S, 2009ApJ...697...16O, 2022ApJ...939...31M}. The resulting disc is, however, geometrically much thicker than any prediction of the standard models allows, and so its observable properties are distinct from the widely-used accretion disc models \citep{2022MNRAS.514..780W}. 

The structure of the disc, as shown in the left panel of Figure \ref{fig:puffy}, can be separated into two optically thick regions -- the geometrically thin dense core (C) and the puffy region (P), surrounded by an optically thin funnel (F) dominated by outflow. The core is defined by the scale-height, which is significantly different from the photosphere height, which encloses the puffy region. The photospheric height of the disc is comparable to the radial distance from the centre ($H/r \sim 1$), whereas the standard disc models predict $H/r < 0.3$ for the same mass accretion rate. The puffy disc structure resembles a thin disc surrounded by a warm corona \citep{2015A&A...580A..77R,2020A&A...633A..35G}. However, the puffy region also supplies most of the mass accretion, in comparison with the very low inflow in the central core.

Figure \ref{fig:puffy} shows the time and azimuth-averaged structure of the puffy disc. The left panel shows the gas density in colours, the magnetic field lines in black contours, and labels of the three regions. The right panel shows the two-dimensional map of $\alpha$, where the increase inside the plunging region is apparent -- the yellow contour shows the magnetosonic surface, that defines the approximate separation between the plunging region and the rotating disc, which we use as the sonic point in this work. For details of the calculations, see \citetalias{lancova_new_2025}.

%
%
\subsection{The puffy discs stress calculations}
\label{sec:puffy-discs-alpha}
\begin{figure} [!h]
\begin{center}
\includegraphics[width=0.98\columnwidth]{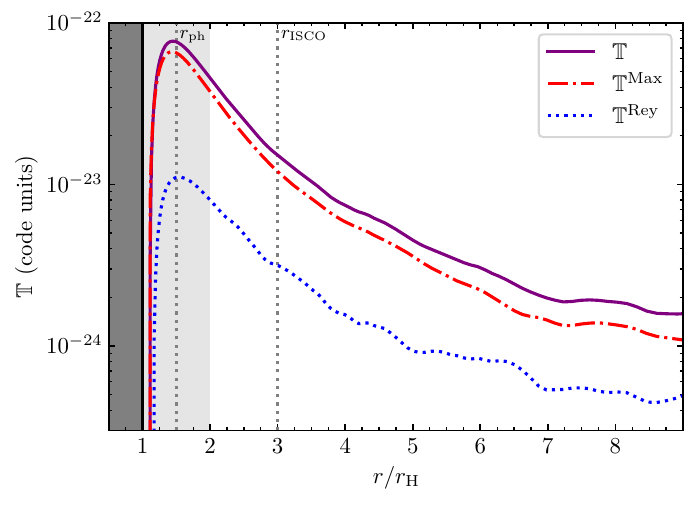}
\caption
{ 
Behaviour of the total stress $\mathbb{T} = \mathbb{T}^{\mathrm{Max}} + 
\mathbb{T}^{\mathrm{Rey}}$ and its Reynolds and Maxwell components (the lines 
shown are taken from  \citetalias{lancova_new_2025}): the stress is zero at the horizon, 
it has a maximum at the circular photon orbit and is much smaller far away than 
in the plunging region. The behaviour of the stress is remarkably 
similar to the behaviour of the inverse of the square of the ``gyration radius'' $\tilde{r}$. 
} 
\label{Fig:three-stresses}
\end{center}
\end{figure}

In extracting $\alpha$ from GRMHD simulations, \citetalias{lancova_new_2025} followed the method described in \citetalias{2013MNRAS.428.2255P} to calculate the stress, which consists of three steps. First one defines the time-averaged three-dimensional four-velocity $\langle u^\mu \rangle_t = U^\mu$. Then, one constructs the co-moving tetrad $e^\nu_{\,\,(k)}$. Subsequently, one calculates the total stress and its Reynolds and Maxwell parts, as the tetrad ${(r)(\phi)}$ components of the corresponding stress-energy tensors computed from individual 3D snapshots and subsequently averaged over time and azimuth,
$\langle T_{(r)(\phi)}^\mathrm{Tot} \rangle_{t,\phi}$,
$\langle T_{(r)(\phi)}^\mathrm{Rey} \rangle_{t,\phi}$,
$\langle T_{(r)(\phi)}^\mathrm{Max} \rangle_{t,\phi}$,
according to equation (\ref{eqn:invariant-stress}). The density-weighted vertically averaged stress and its components are shown in Figure \ref{Fig:three-stresses}. Finally, $\alpha$ is 
\begin{equation}
    \alpha= \frac{\mathbb{T}}{\langle p\rangle_{t,\phi}},
\end{equation}
in both \citetalias{lancova_new_2025} and \citetalias{2013MNRAS.428.2255P}, where $\langle p\rangle_{t,\phi}$ is the averaged total pressure. Then, a radial profile of $\alpha$ is obtained by a density-weighted average of the two-dimensional $\alpha$ map (shown in the right panel of Figure \ref{fig:puffy}) in the vertical direction.

\citetalias{lancova_new_2025} included the gas, magnetic, and radiation pressures for the $\alpha$ calculations, $p = \pgas + \pmag + \prad$. This is in contrast with \citetalias{2013MNRAS.428.2255P}, who used only the magnetic and gas pressures, $p = \pgas + \pmag$, since their simulations do not provide the radiation pressure component. \citetalias{2025MNRAS.542..377R} further used only the instantaneous fluid frame Maxwell stress and gas and magnetic pressure.

%
%
\section{Discussion and future work}
\label{sec:future-work}
In this paper, we report that the results of GRMHD simulations of the black hole accretion disc obtained in several independent studies by different authors seem to indicate a universal radial dependence of the Shakura-Sunyaev $\alpha$-viscosity coefficient, which most probably reflects fundamental properties of the stress itself. This finding suggests two directions of future research:
First, an explanation from first principles of the universal stress behaviour seen in Figure \ref{Fig:three-stresses}, 
and second, an application of our $\alpha$-viscosity formula (\ref{eqn:whole-domain}) to construct semi-analytic models of thin and slim discs.
\subsection{Fundamental stress properties}
\label{sec:fundamental-stress-properties}
We will discuss fundamental physical reasons for such behaviour in a forthcoming, much more mathematically minded publication. In this subsection, we only briefly hint at some relevant points, appealing to physical intuition rather than giving solid mathematical proofs. One of the issues that we will carefully clarify (because it is often a~source of confusion) is that: the stress is here zero at the horizon, even though the equivalence principle assures that one cannot know from local measurements if the horizon is being crossed. In a wider context --- the MRI builds the stress locally, in the comoving frames, but some of the stress properties reflect the global aspects of the black hole space-time.  
\subsubsection{Stress is zero at the horizon}
\label{sec:stress-horizon-zero}
Let us cast equation (\ref{eqn:momentum-energy-flux}) in the form  
\begin{equation}
\mathbb{T}= {\dot J} - \langle{\dot M}\rangle\,\langle j \rangle
\label{eqn:stress-Feynman-definition}
\end{equation}
that underlines the physical meaning of the stress: it represents the part of the angular momentum flux through a surface ${\cal S}$ that is not carried with the mean mass accretion flux $\langle{\dot M}\rangle \langle j \rangle$. Here $\langle j \rangle$ is the mean specific angular momentum at ${\cal S}$. \citet{2010A&A...521A..15A} applied
(\ref{eqn:stress-Feynman-definition}) and argued that when a sphere $\cal{S}$ of radius $r$ in an accretion disc is crossed by a ``mean'' accretion flow $\langle{\dot M}\rangle$ in the $+$ direction (towards the black hole), turbulence causes an additional (fluctuating) downstream flux $\delta{\dot M}^+$ and an upstream flux $\delta{\dot M}^-$, 
and therefore the total fluxes of mass in the $+$ and $-$ directions are  ${\dot M}^+ = \langle{\dot M}\rangle + \delta{\dot M}^+$ and ${\dot M}^- = \delta{\dot M}^-$. Therefore, 
\begin{equation}
\mathbb{T} = \frac{1}{2} (\delta {\dot M}^-)(\delta j),
\label{eq:stress-trough-surface}
\end{equation}
where $(\delta j) = [{\dot M}^+ j^+ - {\dot M}^- j^-]/{\langle {\dot M} \rangle}$. 

From equation (\ref{eq:stress-trough-surface}), \citet{2010A&A...521A..15A} have deduced that, if the surface ${\cal S}$ coincides with the black hole horizon, then no upstream flux of mass exists, $(\delta{\dot M}^-=0)$ and hence, there is no stress at $r = r_{\mathrm{H}}$. This simple argument, which captures the essence of the relevant physics, was later extended by \citet{2014PhRvD..89b4041L} who considered  
an arbitrary stress-energy tensor $T_{\mu \nu}$, that describes {\it any kind} of matter or field, in particular the electromagnetic field. Specifically, for the electromagnetic field in the GRMHD environment, \citet{2014PhRvD..89b4041L} {\it rigorously proved} that the Blandford-Znajek mechanism is an electromagnetic version of the Penrose process. They showed that, although angular momentum is transported along the field lines of a strongly enhanced magnetic field, no magnetic stress (torque) is exerted at the horizon. Instead, the black hole absorbs negative energy and negative angular momentum, resulting in a reduction of its rotational energy. As a particular example, \citet{2014PhRvD..89b4041L} showed that this general principle (``no stress but negative energy absorption'') takes place in magnetically arrested accretion discs \citep{1974Ap&SS..28...45B,2003PASJ...55L..69N}.

The vanishing of the turbulent stress in the mean comoving frame at the horizon, which we use in the definition of viscous $\alpha$, cannot be directly linked only to crossing the horizon: by the equivalence principle, a local comoving observer does not recognise the horizon crossing. Nonetheless, GRMHD simulations show that the flow at the horizon is nearly laminar and almost radial. One reason for this is that once the radial motion becomes highly supersonic and super-Alfvénic, the MRI stops being an efficient driver of turbulence, and both the turbulent Reynolds and Maxwell stresses decay quickly. While the Reynolds stress calculated in the GRMHD simulations discussed here is the turbulent one, the Maxwell stress in the plunging region is mainly due to the large-scale magnetic field. However, when the azimuthal component of the mean magnetic field becomes negligible in the ideal MHD flow due to the field lines being frozen into the radially infalling fluid, the relevant component of the Maxwell stress also vanishes.

\subsubsection{Stress has its maximum at the photon radius}
\label{sec:stress-maximum-photon}   
Figure \ref{Fig:three-stresses} shows that GRMHD simulations clearly indicate that both the Reynolds and Maxwell components of the stress (and therefore also the total stress) peak sharply at the location of the radius of the circular photon orbit. In a forthcoming analytic study we plan to address the question of whether the stress maximum is caused by the reversal of the sense of {\it outward} and {\it inward} radial directions that is known to occur at the location of the circular photon orbit $r_\mathrm{ph}=(3/2)r_{\mathrm{H}}$. For $r<$\,$r_{\mathrm{ph}}$, ``out'' points towards the black hole at the centre \citep{Wex-1995, 1990MNRAS.245..720A}. The reversal influences all of the dynamical properties of rotation, in particular the reversed Rayleigh criterion demands that, for stability, angular momentum should {\it increase inwards}, meaning towards the black hole. Correspondingly, the Rayleigh stress transports angular momentum {\it inwards}, into the black hole \citep{1988MNRAS.233..489A}. There are several other examples of this reversal described in a popular Scientific American article by \cite{1993SciAm.268c..74A}.      The reversal of the dynamical sense of the outward and inward directions could be another reason for decreasing stresses close to the horizon, discussed in the previous subsection. As this purely geometric effect causes a shear reversal, one may speculate that the reversed shear suppresses the MRI and causes the turbulent stress to vanish.

According to our proposed prescription given in equation~(\ref{eqn:whole-domain}), $\alpha(r)$ reaches maximum exactly at the photon orbit. However, our argumentation pertains to the behaviour of stress, rather than $\alpha$ which is proportional to stress and inversely proportional to pressure. The radial behaviour of stress, with a maximum located exactly at the photon orbit is shown in Figure~\ref{Fig:three-stresses}. Then the small shift of the maximum of $\alpha(r)$ away from the photon orbit, as seen in Figure~\ref{fig:universal-stress}, is caused by the radial dependence of pressure. Neglecting this small shift in our prescription and framing the discussion around $\alpha$ is motivated by historical and practical reasons, related to applications for modelling of accretion discs.
\subsubsection{At large radii $\alpha=$\,(shear)/(vorticity) }
\label{sec:stress-large-radii}   
In the far region $r \gg r_{\mathrm{H}}$
\begin{equation}
\alpha \approx \alpha_\infty = \mbox{const}, 
\label{eqn:far-domain}
\end{equation}
with $10^{-2} \lesssim \alpha_{\infty} \lesssim 10^{-1}$. \cite{1996MNRAS.281L..21A} noticed that in the far region, 
for flows that are dominated by matter spiralling down along orbits that may be considered almost circular, a more detailed fitting is consistent with 
\begin{equation}
\alpha \approx \frac{\alpha_\infty}{3} \left( \frac{\sigma}{\omega}\right),
\label{eqn:far-sigma-omega}
\end{equation}
where the kinematic invariants shear\,$=\sigma$ and vorticity\,$=\omega$ are given in the Schwarzschild space-time by
\begin{align}
\sigma^2 = - \frac{1}{4} \left(1 -\Omega j\right)^{-2} & ({\tilde r})^{+2} (\nabla^\mu \Omega)(\nabla_\mu \Omega), \\
\omega^2 = - \frac{1}{4} \left(1 -\Omega j\right)^{-2} & ({\tilde r})^{-2} (\nabla^\mu j)(\nabla_\mu j).
\end{align}
This possibly reflects the origin of the stress as driven by the {\it local} MRI, with turbulence being sensitive to local gradients of angular velocity $\Omega$ and specific angular momentum $j$.

The numerical results of \cite{1996MNRAS.281L..21A} for stratified MRI turbulence were verified by
\cite{1999ApJ...518..394H}, who considered unstratified MRI turbulence for a~fine mesh of shear parameters $q$.
In their local shearing box simulations, $q=-d\ln\Omega/d\ln r$ is the double-logarithmic angular velocity gradient,
which quantifies the strain of the background velocity  $\Omega_0$, $\sigma=q\Omega_0/\sqrt{2}$.
By contrast, the vorticity of the background velocity is $\omega=r^{-1}d(r^2\Omega/dr)/\sqrt{2}=(2-q)\Omega_0/\sqrt{2}$.
\cite{1999ApJ...518..394H} noted that for small values of $q$, the turbulent fluctuations in the stress are dominated by the magnetic contributions.
Interestingly, based on their analytical considerations, they also pointed out that the magnetic stress fluctuations couple directly to the strain $\sigma\propto q$,
while the kinetic stress fluctuations couple to the vorticity $\omega\propto2-q$.
Therefore, $\sigma/\omega=q/(2-q)$, so for small values of $q$, the vorticity $\omega\propto2-q$ limits the turbulent transport quantified by $\alpha$, while strain $\sigma\propto q$ promotes it.

These results were subsequently corroborated by \cite{2006PhRvL..97v1103P, 2008MNRAS.383..683P} using a linear stability analysis.
In particular, they confirmed the observation that the turbulent stresses are not simply proportional to the local shear, as one might have naively expected.
Similar conclusions were also drawn by \cite{2009AN....330...92L} using a closure model of \cite{2003MNRAS.340..969O}.
The numerical results of \cite{1996MNRAS.281L..21A} and \cite{1999ApJ...518..394H} were later also confirmed for much taller shearing boxes of up to eight scale heights
\citep{2015MNRAS.446.2102N}.
%
%
\section{Conclusions}
\label{sec:conclusions}

We have suggested an improvement to the well-known, analytic, Shakura-Sunyaev $\alpha$-prescription for the stress $\mathbb{T} = \alpha p$. Instead of the original ansatz $\alpha = \rm const$, we propose
\begin{equation}
\alpha = \alpha_P \left({r_{\mathrm{H}}}/{{\tilde r}}\right)^2 + \alpha_{\infty}\left(\eta\eta\right), 
\quad {\tilde r}^2 = - \frac{{(\xi \xi)}}{(\eta \eta)},
\label{eqn:alpha-conclusions}
\end{equation} 
with $\alpha_P \sim 1$, $\alpha_{\infty} \sim 0.01$ being phenomenological constants and $r_{\mathrm{H}}=2\mathrm{G}M/\mathrm{c}^2$ being the non-rotating black hole horizon radius.

The values of $\alpha_P$ and $\alpha_\infty$ can be connected to the values used in standard disc modelling (e.g. $\alpha_\infty \sim 0.1$), and to the magnetisation of the matter in the innermost region, e.g., through scaling $\alpha = 0.5\beta^{-1}$, where $\beta$ is the ratio of thermal to magnetic pressure \citep{sorathiaGlobalSimulationsAccretion2012}. A more detailed investigation of how disc properties and magnetisation affect the shape of the $\alpha(r)$ profile is left for future work. However, in \citetalias{lancova_new_2025} we showed that the $\alpha$ behaviour does not vary significantly for simulations covering the mass accretion rate range of $\dot{m} = 0.4$--$0.9\,\dot{M}_\mathrm{Edd}$. Simulations by \citetalias{2013MNRAS.428.2255P} and \citetalias{2025MNRAS.542..377R} were performed for lower mass accretion rates and lower disc magnetisation and show lower values of the peak $\alpha$ (see the caption of Figure~\ref{fig:universal-stress}), but our formula fits those results well with different values of the $\alpha_P$ and $\alpha_\infty$ parameters.

The proposed analytic formula is based on the results of GRMHD simulations that describe black hole accretion discs with sub-Eddington and around-Eddington accretion rates. It is analytic and expressed in terms of coordinate-independent quantities --- the Kerr metric Killing vectors $\eta^\mu$, $\xi^\mu$. It is short, simple, and explicit in any specific coordinates used in the black hole accretion studies. It accurately embraces fundamental properties of the stress in the curved black hole space-time: the stress is zero at the event horizon, it is smaller far away from the black hole than in the plunging region closer to it, and it peaks near the location of the circular light radius.

We plan to use (\ref{eqn:alpha-conclusions}) to calculate improved models of both stationary and non-stationary slim accretion discs. For the stationary ones, we will first need to convert the equations given by \cite{2009ApJS..183..171S} into the comoving frame so as to ensure that the stress description (\ref{eqn:alpha-conclusions}) corresponds directly to that used in \citetalias{lancova_new_2025} and other GRMHD simulations. For the non-stationary ones, the simulations by \cite{1998MNRAS.298..888S} and \cite{2001MNRAS.328...36S} have already been performed in a Lagrangian frame, as is optimal for various reasons. Further work will involve applying (\ref{eqn:alpha-conclusions}) into a similar scheme but with modifications to give an improved relativistic treatment of the accretion flow between the sonic radius and the black hole horizon, enabling it to follow the essential characteristics of the flow there. This approach should make it possible to calculate a long time evolution of the disc, as before, which is crucial for understanding the observational appearance of objects powered by accretion onto black holes. 

The most exciting research outcome that one may anticipate from the findings of this paper is the possibility of explaining these features of the stress:
\begin{equation*}
\mathbb{T}(r_\mathrm{H})=0, ~\,~ \mathbb{T}(r_{\rm ph})=\mathbb{T}_{\rm max}, ~\,~ \mathbb{T}(r\gg r_\mathrm{H}) \propto \sigma/\omega,   
\end{equation*} 
which may reflect the way in which the quasi-local MRI turbulence interacts with the global constraints coming from the space-time geometry of the black hole.

This paper is an initial attempt to extract useful information about the stress in accretion discs around black holes from the numerical results of recent GRMHD simulations. We point out that the radial stress profiles from three independent and quite different numerical simulations return a very similar profile, and this is our motivation for proposing its ``universal'' character: (1) the stress measured in the comoving frame is zero at the black hole horizon and (2) it has a maximum at the location of the circular photon orbit.

Of course, as these statements are based on numerical results, one may have doubts about whether the zero point of the stress really occurs exactly at the horizon. It would be very desirable to have an analytic investigation of this --- and we are working on that. The same concerns the need to have equivalent results for rotating black holes in the hope of finding a more universal solution.

There are currently very few available GRMHD simulations of accretion onto a spinning Kerr black hole in a regime where the $\alpha$-viscosity prescription is applicable in analytical modelling, and even fewer radiative GRMHD simulations where the results are not affected by the choice of cooling function or disc instabilities, and where it is possible to study the flow properties in the innermost regions around the black hole, \citep[e.g.,][and others]{2023ApJ...959...59F,2026arXiv260305588Z}. We aim to explore this regime with puffy disc simulations in the future. In any case, exploring the parameter space of various black hole masses, spins, and mass accretion rates would be extremely computationally expensive, and hence analytic studies are very much needed here.

The most obvious astrophysical applications of our paper are to construct analytic and semi-analytic models of accretion with the aid of the new viscosity prescription, particularly for slim discs. However, these studies may give results that would be far more general than the astrophysical accretion disc theory itself: for example, that the stress profile, independently of the nature of the stress, is shaped near to the horizon mainly by gravity, and not by properties of the fluid.


\begin{acknowledgements}
We are grateful to the anonymous referee for valuable suggestions. The first version of this draft was prepared at the RAGtime workshop in Opava. MA and JH acknowledge the Czech Science Foundation (GA\v{C}R) grant No. 21-06825X, and DL project No. 25-16928O. AB acknowledges support by the Swedish Research Council (Vetenskapsr{\aa}det) under grant No.\ 2025-05957. JM acknowledges continuing support from the Italian Istituto Nazionale Previdenza Sociale and provision of facilities by the University of Oxford. MW is supported by a Ramón y Cajal grant RYC2023-042988-I from the Spanish Ministry of Science and Innovation and acknowledges financial support from the Severo Ochoa grant CEX2021-001131-S funded by MCIN/AEI/ 10.13039/501100011033. This work was supported by the Ministry of Education, Youth and Sports of the Czech Republic through the e-INFRA CZ (ID:90254).\\

\textit{Data availability.} The simulation data underlying this article are available upon reasonable request to the corresponding authors.
\end{acknowledgements}
\bibliographystyle{aa} 
\bibliography{bibliography} 
\appendix

\section{Kerr metric formulae}
\label{app:Kerr-formulae}
\noindent
The covariant definitions of the quantities used in the paper are given in terms of Kerr metric symmetry generators: the Killing vector of time symmetry $\eta^i$ and the Killing vector of axial symmetry $\xi^i$. The Killing vectors obey
\begin{equation}
\nabla_{(i}\eta_{k)} = 0, \quad \nabla_{(i}\xi_{k)} = 0, \quad \eta^i \nabla_i \xi_k = \xi^i \nabla_i \eta_k.
\label{eq:app-Killings}
\end{equation}   
The frame dragging $\omega$ and the gravity potential $\Phi$ obey
\begin{equation}
\omega = - \frac{(\eta \xi)}{(\xi \xi)}, \quad 
\Phi = -\frac{1}{2} \ln \bigl[ (\eta \eta) + 2\omega\,(\eta \xi) + \omega^2 (\xi \xi) \bigr]
\label{eq:app-dragging-and-gravity}
\end{equation}   
which help to define the four-velocity of the stationary observer (ZAMO) 
\begin{equation}
n^i = e^{\Phi} (\eta^i + \omega \xi^i), \quad (nn) = 1.
\label{eq:app-ZAMO}
\end{equation}   
A circular motion at the equatorial plane symmetry of the Kerr metric is defined by writing two equivalent expressions for the four-velocity of matter,
\begin{align}
u^i &= A (\eta^i + \Omega \xi^i), \quad A^{-2} = (\eta \eta) + 2\Omega\,(\eta \xi) + \Omega^2 (\xi \xi),
\label{eq:app-omega-02} \\
u^i &= \gamma (n^i + v \tau^i), \quad \tau^i = \xi^i/r, \quad r^2 = -(\xi \xi).
\label{eq:app-four-velocity-02}
\end{align}
These equations invariantly define the angular velocity $\Omega$, the orbital velocity $v$ and the redshift factor $\gamma^{-2}=1 - v^2$. 
The specific angular momentum $j$ obeys
\begin{equation}
j = - \frac{(u\xi)}{(u\eta)}, \quad j = v {\tilde r} = {\tilde \Omega} {\tilde r}^2,
\label{eq:app-specific-momentum}
\end{equation}   
where ${\tilde \Omega} = \Omega - \omega$ and
\begin{equation}
{\tilde r} = r e^{\Phi}
\label{eq:app-gyration}
\end{equation}   
is the gyration radius; see \cite{Wex-1995} for more details.

                                                                                      \end{document}